\documentclass[preprint]{aastex}
\renewcommand\bv{{\bf v}}
\renewcommand\>{{\rangle}}
\newcommand\bnabla{{\bf \nabla}}
\newcommand\bk{{\bf k}}
\newcommand\br{{\bf r}}
\newcommand\bx{{\bf x}}
\newcommand\ex{\hat{\bf e}_x}
\newcommand\ey{\hat{\bf e}_y}
\newcommand\tc{{\tau_{c}}}
\newcommand\bO{{\bf \Omega}}
\newcommand\del{\partial}
\newcommand\<{{\langle}}

\newcommand\eps{{\epsilon}}

\newcommand\qq{{3\over{2}}}

\newcommand\au{{\rm\,AU}}

\newcommand\yr{{\rm\,yr}}

\newcommand\kms{{\rm\,km\,s^{-1}}}

\newcommand\msun{{\rm\,M_\odot}}

\newcommand\K{{\rm\,K}}
\shortauthors{Gammie}
\shorttitle{Gravitational Instability}
\begin{document}

\title{Nonlinear Outcome of Gravitational Instability in Cooling,
Gaseous Disks}

\author{Charles F. Gammie}

\affil{Center for Theoretical Astrophysics,}
\affil{Physics Department, and Astronomy Department}
\affil{University of Illinois at Urbana-Champaign}
\affil{1110 W. Green St.}
\affil{Urbana, IL 61801}

\begin{abstract}

Thin, Keplerian accretion disks generically become gravitationally
unstable at large radius.  I investigate the nonlinear outcome of such
instability in cool disks using razor-thin, local, numerical models.
Cooling, characterized by a constant cooling time $\tc$, drives the
instability.

I show analytically that, if the disk can reach a steady state in which
heating by dissipation of turbulence balances cooling, then the
dimensionless angular momentum flux density $\alpha = \left((9/4) \gamma
(\gamma-1) \Omega \tc\right)^{-1}$.  Numerical experiments show that:
(1) if $\tc \gtrsim 3\Omega^{-1}$ then the disk reaches a steady,
gravito-turbulent state in which $Q \sim 1$ and cooling is balanced by
heating due to dissipation of turbulence; (2) if $\tc \lesssim
3\Omega^{-1}$, then the disk fragments, possibly forming planets or
stars; (3) in a steady, gravito-turbulent state, surface density
structures have a characteristic physical scale $\sim 64 G
\Sigma/\Omega^2$ that is independent of the size of the computational
domain.

\end{abstract}

\keywords{accretion, accretion disks, solar system: formation, galaxies:
nuclei}

\section{Introduction}

It has long been realized that the outer reaches of accretion disks
around active galactic nuclei (AGN) and young stellar objects (YSO) may
be gravitationally unstable (for a review see, for AGN: \cite{shlos90};
YSOs: \cite{al93}).  Instability in a Keplerian disk sets in where the
sound speed $c_s$, the rotation frequency $\Omega$, and the surface
density $\Sigma$ satisfy
\begin{equation}\label{QDEF}
Q \equiv {c_s\Omega\over{\pi G\Sigma}} < Q_{crit} \simeq 1
\end{equation}
\citep{toom64,glb}.  Here $Q_{crit} = 1$ for a ``razor-thin'' (two
dimensional) fluid disk, and $Q_{crit} = 0.676$ for a finite thickness
isothermal disk \citep{glb}.  The instability condition (\ref{QDEF})
can be rewritten, for a disk with scale height $H \simeq c_s/\Omega$,
around a central object of mass $M_*$,
\begin{equation}
M_{disk} \gtrsim {H\over{r}} M_*,
\end{equation}
where $M_{disk} \equiv \pi r^2 \Sigma$.  

In a steady-state disk whose heating is dominated by interior turbulent
dissipation and whose evolution is controlled by internal transport of
angular momentum, the accretion rate $\dot{M} = 3\pi \alpha c_s^2
\Sigma/\Omega$, where I have used the $\alpha$ formalism of \cite{ss73}.
Then 
\begin{equation}\label{INSTCRIT}
\dot{M} \gtrsim {3 \alpha c_s^3\over{G}}
= 7.1 \times 10^{-4}\, \alpha 
\left( {c_s \over{ 1\kms}}\right)^3 \,\msun \yr^{-1}
\end{equation}
implies gravitational instability (e.g. \cite{shlos90}).  This does not
apply to disks dominated by external torques (e.g., a
magnetohydrodynamic [MHD] wind) or disks heated mainly by external
illumination.  

For a young, solar mass star accreting from a disk with $\alpha =
10^{-2} $ at $10^{-6} \msun \yr^{-1}$ equation (\ref{INSTCRIT}) implies
that instability occurs where the temperature drops below $17 \K$.
Disks may not be this cold if the star is located in a warm molecular
cloud where the ambient temperature is greater than $17 \K$, or if the
disk is bathed in scattered infrared light from the central star
(although there is some evidence for such low temperatures in the solar
nebula, e.g. \cite{owen99}).  If the effective value of $\alpha$ is
small and heating is confined to surface layers, however, as in the
layered accretion model of \cite{gam96a}, then instability can occur at
much higher temperatures.

AGN disk heating is typically dominated by illumination from a central
source.  The temperature then depends on the shape of the disk.  If the
disk is flat or shadowed, however, and transport is dominated by
internal torques, one can apply equation (\ref{INSTCRIT}).  For example,
in the nucleus of NGC 4258 \citep{miy95} the accretion rate may be as
large as $10^{-2} \msun \yr^{-1}$ \citep{las96,gam99}.  Equation
(\ref{INSTCRIT}) then implies that instability sets in where $T < 10^4
(\alpha/10^{-2}) \K$.  If the disk is illumination-dominated, however,
then $Q$ fluctuates with the luminosity of the central source.

The fate of a gravitationally unstable YSO or AGN disk depends on how it
arrived in an unstable state.  To understand why, consider an analogy
with convective instability in stellar structure theory.  The evolution
of stellar models with highly unstable radial entropy profiles are of
little interest, because convection prevents such models from being
realized in an astrophysically plausible setting.  Similarly, highly
unstable disks may be irrelevant because the action of the instability
would prevent one from ever arriving in such a state.  Gravitational
instability must be ``turned on'' in a natural way.

An initially stable Keplerian disk can be driven unstable by an increase
in surface density (e.g. \cite{sc84}) or by cooling.  In an $\alpha$
disk model, the surface density changes on the accretion timescale $\sim
(r/H)^2 (\alpha\Omega)^{-1}$  (more rapid variation of surface density
could be obtained by dumping material onto the disk or by application of
a direct magnetic torque).  The temperature, by contrast, changes on the
cooling time, $\sim (\alpha \Omega)^{-1}$.  This suggests that in cool
disks ($r/H \gg 1$), cooling is the dominant driver of gravitational
instability.

Once gravitational instability sets in, the disk can attempt to regain
stability by rearranging its mass to reduce $\Sigma$, or by heating
itself through dissipation of turbulence.  Dissipation can occur
directly through shocks or indirectly via a turbulent cascade to the
viscous scale.  If one can model the effects of ``gravito-turbulence''
on a cool disk as an $\alpha$ viscosity, then mass shifting can be
accomplished only on the accretion timescale $(r/H)^2 (\alpha
\Omega)^{-1}$.  Heating occurs on the thermal timescale $(\alpha
\Omega)^{-1}$.  This suggests that cool disks will attempt to
reestablish stability through dissipation of turbulence.

Let us assume that cooling drives the disk toward instability, and that
the disk tries to recover through turbulent dissipation.  There is now
the possibility of a feedback loop.  If $Q$ is too large then heating by
gravito-turbulence is weak and the disk cools toward instability.  If
$Q$ is too small, heating by gravito-turbulence is strong and the disk
is pushed back toward marginal stability.  The feedback loop acts as a
thermostat to maintain $Q \sim 1$, just as convection in stars drives
the entropy profile toward marginal stability.

The effectiveness of the feedback loop in maintaining $Q \sim 1$ is
controversial.  \cite{lp87,lp90} have considered disk models with $Q \ll
1$.  Indeed it is not clear that the feedback is strong enough to
preserve $Q \sim 1$.  Cooling might be so rapid that the disk
fragments before it can heat \citep{sb89}, leading to formation of gas
giant planets or stars.

The feedback loop is an old idea in the theory of galactic disks
(\cite{glb}; there heating due to star formation balances cooling) and
in accretion disk theory \citep{pac78}.  Numerical experiments on
cooled collisionless disk models \citep{ac88,cf85,tom91,tom94} do yield
disks with $Q \sim 1$, and clumping with sufficiently strong cooling.
Observations of star forming regions of galactic disks also show $Q
\sim 1$ in the gas \citep{ken89}.

Numerical experiments with fluid disks often assume a fixed temperature
profile or a fixed entropy profile \citep{bos97, bos98,nel98}.  A
fixed temperature profile is relevant to an illumination-dominated disk.
Other work considers the adiabatic evolution of an initially unstable
state \citep{pick00,lau96,yan91}.  Recent work by \cite{nel00}
is most comparable to that presented here.  It examines the evolution of
cooling, global models of the disks, but with a more complicated
treatment of the cooling function.  

In this paper I consider the nonlinear evolution of cooling,
self-gravitating fluid disks via numerical experiment.  The disk model
is razor-thin (two dimensional) and local.  In \S 2 I describe the
model, then show that its angular momentum flux density can be
calculated analytically (\S 3), provided that the disk does not
fragment.  The results of numerical experiments are described in \S 4.
Conclusions are given in \S 5.

\section{Model}

The direct, numerical evolution of a global self-gravitating disk model
over many dynamical times is prohibitively expensive.  To develop a
converged, calculable, numerical model, and to reduce the number of free
parameters associated with a global disk model, I make two simplifying
assumptions: that the disk is cool (i.e. thin, so $H/r \simeq
c_s/(\Omega r) \ll 1$) and that it razor-thin. 

A cool disk can be modeled using a rigorous expansion of the equations
of motion through lowest order in $c_s/(\Omega r)$.  This model is
called the ``local model'' or ``shearing sheet.''  The expansion
proceeds as follows.  Choose a fiducial point $(r_o,\phi_o + \Omega(r_o)
t)$ that corotates with the disk.  Define a set of local Cartesian
coordinates $x,y \equiv r - r_o, r_o (\phi - \phi_o - \Omega(r_o) t)$,
and expand the equations of motion to first order in $|{\bf x}|/r_o$.
To do this, one needs to assume that the departures from circular orbits
in the disk $|\delta \bv| \sim c_s \ll \Omega r_0$, and similarly that
the perturbed potential $\delta\phi \sim c_s^2 \ll (\Omega r_0)^2$.
With this assumption, the local model describes the evolution of the
disk near the fiducial point.  The self-consistency of these orderings
can be tested in the nonlinear evolution.

I will also use a ``razor-thin'' disk approximation that treats the disk
as two dimensional.  There is good agreement between razor-thin models
and full three dimensional evolution in the linear theory of polytropic
slender tori (Goldreich et al. 1986), in nonlinear simulations of
polytropic tori (Hawley 1990), and in nonlinear simulations of gas flow
in spiral arms with an isothermal equation of state (Tubbs 1980).  In
general there can be no rigorous mapping between razor-thin disks and
full three-dimensional systems because the vertical structure contains
degrees of freedom that are absent in the razor-thin model.  

The equations of motion in the local model read, where $\bv$ is the
velocity, $P$ is the (two-dimensional) pressure, and $\phi$ is the
gravitational potential with the time-steady axisymmetric component
removed:
\begin{equation}
{D\bv\over{D t}} = -{\bnabla P\over{\Sigma}} - 2\bO\times\bv
        + 3\Omega^2 x \ex - \bnabla\phi.
\end{equation}
For constant pressure and surface density, $\bv = -{3\over{2}}\Omega x
\ey$ is an equilibrium solution to the equations of motion.  This linear
shear flow is the manifestation of differential rotation in the local
model. 

The equation of state is 
\begin{equation}
P = (\gamma - 1) U,
\end{equation}
where $P$ is the two-dimensional pressure and $U$ the two-dimensional
internal energy.  I will adopt $\gamma = 2$ throughout.  The two
dimensional (2D) adiabatic index $\gamma$ can be mapped to a 3D
adiabatic index $\Gamma$ in the low-frequency (static) limit.  For a
nonself-gravitating disk $\gamma = (3\Gamma - 1)/(\Gamma + 1)$ (e.g.
Goldreich et al. 1986, Ostriker et al. 1992).  For a strongly
self-gravitating disk, one can show that $\gamma = 3 - 2/\Gamma$.

If the evolution is strictly adiabatic and the initial conditions are
isentropic, then $p = K \Sigma^\gamma$ with $K = const.$  It follows
that the ``potential vorticity''
\begin{equation}\label{POTVORT}
\xi \equiv {\bnabla \times \bv + 2 \Omega\over{\Sigma}}
\end{equation}
obeys $D \xi/D t = 0$.  There is a close connection between the
conservation of potential vorticity and the stabilizing effects of
rotation \citep{lb66,hunt64}.  Processes that cause the potential
vorticity to evolve can compromise rotational support of the disk at
long wavelengths \citep{gam96b}.

The internal energy equation is
\begin{equation}
{\del U\over{\del t}} + \nabla \cdot (U \bv) = 
	-P\bnabla\cdot\bv - {U\over{\tc}}
\end{equation}
The final term is the cooling function, and $\tc$ is the cooling time.
In general $\tc = \tc(\Sigma,U,\Omega)$.  I will make the simplifying
assumption that $\tc = const.$  Notice that there is no heating term;
heating is due solely to shocks.  Numerically, the fluid is heated by
artificial viscosity in shocks.

The gravitational potential is determined by the razor-thin disk Poisson
equation:
\begin{equation}
\nabla^2\phi = 4\pi G \Sigma \, \delta(z).
\end{equation}
For a single Fourier component of the surface density $\Sigma_{\bk}$
this has the solution
\begin{equation}
\phi = -{2\pi G\over{|\bk|}} \Sigma_{\bk} e^{i \bk\cdot{\bf x}
	- |k z|}.
\end{equation}
A finite thickness disk has weaker self-gravity, but this does not
qualitatively change the dynamics of the disk in linear theory
\citep{glb}.

\subsection{Boundary Conditions, Numerical Methods, and Tests}

I use the ``shearing box'' boundary conditions, described in detail by
\cite{hgb1}.  They apply to a rectangular domain of size $L_x$ by $L_y$,
although I will assume, unless stated otherwise, that $L_x = L_y \equiv
L$.  The boundary conditions may be written, for dependent variable $f =
(\Sigma, U, v_x, \delta v_y \equiv v_y + {3\over{2}} \Omega x)$,
\begin{equation}
f(x,y) = f(x,y + L)
\end{equation}
\begin{equation}
f(x,y) = f(x + L,y - \qq\Omega L t)
\end{equation}
These boundary conditions have been used to study accretion disks
\citep{hgb1}, galactic disks \citep{tk91}, and planetary rings
\citep{wt88}.

I integrate the governing equations using a self-gravitating
hydrodynamics code based on ZEUS (Stone \& Norman 1992).  ZEUS is a
time-explicit, operator-split, finite-difference method on a staggered
mesh.  It uses an artificial viscosity to capture shocks.  My
implementation has been tested on standard linear and nonlinear
problems, such as sound waves and shock tubes, with acceptable results.

The transport scheme differs significantly from the basic ZEUS
algorithm.  I divide the transport step into two pieces: one due to the
mean velocity $-\qq\Omega x \ey$ and the other due to the remaining,
perturbed velocity.  The mean velocity transport is done by linear
interpolation, while the perturbed velocity transport is done as in
ZEUS.  \footnote{I thank J. Goodman for suggesting this.  See also
\cite{mass99}.} The ``ghost zones'' outside the boundaries are likewise
set using linear interpolation.  This algorithm has the advantage that
the timestep is limited not by the large shear velocity at the radial
edges of the grid, $-\qq\Omega (L/2)$, but by the perturbed velocities
only.  Longer timesteps can therefore be used, and numerical diffusion
is reduced and more nearly uniform across the grid.

I solve the Poisson equation using the Fourier transform method,
modified for the shearing box boundary conditions.  The density is
periodic in shearing coordinates $x,y'$, where
\begin{equation}
y' = y - \qq\Omega x (t - t_p)
\end{equation}
and $t_p = (\qq\Omega)^{-1} {\rm NINT}(t \qq\Omega)$.  Thus when $t =
t_p$, which happens at intervals $\Delta t = (\qq\Omega)^{-1}$, the
model is periodic in $x,y$.  I linear interpolate $\Sigma$ onto a grid
in $x,y'$, solve for the potential via Fourier transform, linear
interpolate the potential back onto a grid in $x,y$, and then difference
to obtain the acceleration.

There is one subtlety involved in the solution of the Poisson equation.
In the Fourier domain, the grid of available wavenumbers forms a
parallelogram in the $k_x,k_y$ plane whose shape changes with time.  I
use only those wavenumbers with $|\bk| < (2)^{-1/2} (N/2) (2\pi/L)$ in
the solution, where $N \times N$ is the numerical resolution of the
model.  This restricts the gravitational kernel to the largest circular
region in Fourier space that is always available.  This restriction
ensures that the gravitational force is approximately isotropic on small
scales.

I have tested my algorithm on three problems.  First, I tested the
transport piece of the code.  I evolved, for one rotation period,
initial conditions in which $v_{x}(t = 0) = const.$, $P = 0$, and a
square of enhanced density was placed in the middle of the grid.  The
square sheared out, crossed a radial boundary, and returned to its
initial position.  The density enhancement diffused somewhat, but no
features were introduced at the boundary crossing.  Second, I tested the
code's ability to conserve potential vorticity.  I evolved initial
conditions containing a sinusoidal velocity perturbation for several
shear times.  Only modest diffusion of potential vorticity was observed.
Finally, I checked the evolution of a sinusoidal density perturbation
against linear theory.

A comparison of code output with linear theory is shown in Figure 1.
The test model has $L = 40 G \Sigma_0/\Omega^2$, $Q = 1$, and no
cooling.  The initial conditions are $\bv = -\qq \Omega x \ey$ and
$\Sigma = \Sigma_0 + \delta \Sigma \cos(\bk \cdot \bx)$, with $k_x = -2
(2 \pi/L)$, $k_y = (2 \pi/L)$, and $\delta\Sigma/\Sigma_0 = 5 \times
10^{-4}$.  Linear theory (see \cite{gam96b}, and references therein)
gives the evolution of the amplitude $\delta\Sigma(t)$ of a shearing
wave $\cos(\bk \cdot \bx)$ where $k_x(t) = k_x(0) + \qq \Omega k_y t$
and $k_y = const.$.  The heavy solid line in the figure shows the linear
theory result.  The other lines show the evolution of the test models,
which start at $N = 32$ (poorest agreement with linear theory) and run
by factors of $2$ up to $N = 256$.  Since the radial wavenumber of the
disturbance increases with time, the wave eventually becomes wrapped up
and unresolved.  At the end of the run ($t = 10\Omega^{-1}$) the radial
wavelength is $L/13$, or $2.46 \times (N/32)$ zones.  Evidently the
numerical integration converges on the linear theory result.

\section{Analytical Properties of the Model}\label{ANALYT}

Consider the following Jacobi-like ``integral'' for the shearing box:
\begin{equation}
\Gamma = \int d^3 {\bf x} \,\,\Sigma \delta(z) \,\, \left(
        {1\over{2}}v^2 + {U\over{\Sigma}} + \phi_T +
        {1\over{2}}\phi\right).
\end{equation}

Here $\phi_T = -\qq\Omega^2 x^2$ is the tidal expansion of the effective
potential about the fiducial point, and $\phi$ is the potential of
surface density fluctuations.  The integral is taken over the entire
simulation area and from $z = -\infty$ to $z = \infty$.  Now evaluate
$\del\Gamma/\del t$ using the energy equation, the equations of motion,
the Poisson equation, and the boundary conditions:
\begin{equation}\label{DGAMMA}
{\partial \Gamma\over{\partial t}} =
\qq\Omega L \int_X dS \left(\Sigma v_x \delta v_y
        + {g_x g_y\over{4\pi G}}\right)
- {1\over{\tc}}\int d^3{\bf x}\,\, U \, \delta(z),
\end{equation}
where the surface integral is taken over one of the radial boundaries
(the azimuthal boundaries do not contribute because they are periodic).
To obtain this result, one must realize that terms proportional to
$D\phi/D t$ vanish when integrated over the boundaries, while terms
proportional to $\del \phi/\del t = D\phi/D t - ({\bf v}\cdot \nabla)
\phi$ do not.  

Now average equation (\ref{DGAMMA}) over $t$, $x$, and $y$, replacing
$\int_X dS f$  by $L\<\int dz f\>$.  If the disk can settle into a
steady state then $\< \del\Gamma/\del_t \> = 0$ and
\begin{equation}\label{AVGSOL}
\left< \Sigma v_x \delta v_y
        + \int \, dz\, {g_x g_y\over{4 \pi G}} \right>
        = {\< U \>\over{\qq \Omega \tc}},
\end{equation}
The left hand side is the shear stress, which is proportional to the
angular momentum flux density.  

The shear stress can be divided into a hydrodynamic piece and a
gravitational piece.  The hydrodynamic shear stress is obtained via the
usual Reynolds decomposition: the fluid velocity is is reduced to the
sum of a mean steady part $-\qq\Omega x \ey$ and a fluctuating part $v_x
\ex + \delta v_y \ey$.  Then the associated ``Reynolds stress'' is
\begin{equation}
H_{xy} = \Sigma v_x \delta v_y.
\end{equation}
Notice that the mean radial velocity is of order $c_s (H/r)^2$, so in the
local model, accurate to first order in $H/r$, the mean radial velocity
is zero.  Thus the local model can be used to calculate a shear stress,
but a global model is needed to relate the shear stress to an accretion
rate.

The gravitational piece of the shear stress is
\begin{equation}
G_{xy} = \int_{-\infty}^{\infty}\,dz\,\, {g_x g_y\over{4\pi G}}
\end{equation}
\citep{lbk}.  The vertical integral arises because the gravitational
field outside the disk contributes to the shear stress.  The vertical
integral can be done analytically in the Fourier domain, giving the
space-averaged gravitational shear stress:
\begin{equation}
\< G_{xy}\> = \sum_\bk \, {\pi G k_x k_y |\Sigma_{\bk}|^2\over
        {|\bk|^3}}.
\end{equation}
The sum is over all Fourier components.

It is convenient to express the shear stress in terms of an effective
$\alpha$:
\begin{equation}
\alpha \equiv {2\over{3 \<\Sigma c_s^2\>}} (G_{xy} + H_{xy}).
\end{equation}
Using equation (\ref{AVGSOL}), and $U = c_s^2\Sigma/(\gamma (\gamma -
1))$, I find
\begin{equation}\label{ANALPHA}
\alpha = \left(\gamma (\gamma - 1) {9\over{4}} \Omega \tc\right)^{-1}.
\end{equation}
For $\gamma = 2$,
\begin{equation}
\alpha = {2\over{9}}{1\over{\Omega \tc}}.
\end{equation}
It is remarkable that it is possible to calculate $\alpha$ analytically
for this simple model.  Equation (\ref{ANALPHA}) does not close the
system of equations for disk evolution, however, because in general
$\tau_c$ is a function of the disk temperature.   It is the local analog
of the result that the surface brightness of the disk is independent of
the details of the angular momentum transport mechanism.  Fixing $Q$
would close the system of equations, but $Q$ is not known {\it a
priori}.  This result, however, can serve as a check on numerical
work.  It also explains the empirical relation between cooling rate and
angular momentum flux reported by \cite{tom94}.

\section{Nonlinear Outcome}

\subsection{Standard Run}\label{STDRUN}

Consider the evolution of a single ``standard'' run, with $L = 320
G\Sigma/\Omega^2$ and $N = 1024$.  The initial velocities are $v_x =
\delta v_x$, $v_y = - {3\over{2}}\Omega x + \delta v_y$, where $\delta
\bv$ is a white noise random velocity field of subsonic amplitude (the
outcome is nearly independent of the details of these perturbations).
The initial internal energy is such that $Q = 1$, and the cooling time
is $\tc = 10\Omega^{-1}$.  

The evolution of the kinetic, gravitational, and thermal energy per unit
area, normalized to $G^2\Sigma^3/\Omega^2$, are shown in Figure 2.  The
evolution of the gravitational and hydrodynamic components of $\alpha$
are shown in Figure 3.  The $\alpha$'s have been smoothed to make the
plot readable.  A snapshot of the surface density at $t = 50\Omega^{-1}$
is shown as a color plot in Figure 4; red is high density and blue is
low density.

As the model evolves the small initial velocity fluctuations grow
exponentially and develop into nonlinear surface density, velocity, and
potential fluctuations.  Shocks of a predominantly trailing orientation
develop and heat the gas, while cooling works to reduce the gas entropy.
Density structures develop which are sheared out by differential
rotation and are also predominantly trailing.  The tendency of velocity
and density structures to take on a trailing figure gives rise to a
finite $\alpha$.

Eventually the thermal energy of the disk settles down to a steady state
in which the disk is near the point of marginal stability, as
hypothesized by \cite{pac78}.  The mean stability parameter $\<Q\>
\equiv \langle c_s \rangle\Omega/\pi G\langle \Sigma\rangle$ fluctuates
slightly, but averages $2.46$ over the last $80 \Omega^{-1}$ of the run.
This is larger than the neutrally stable value ($Q = 1$), but small
enough that the disk remains sensitive to nonaxisymmetric disturbances.
This confirms the hypothesis that the disk maintains $Q \sim 1$.

The analytic theory of the last section predicts that $\alpha = 2/90
\approx 0.022$.  Averaging over the final $80 \Omega^{-1}$ of the run, I
find $\alpha = 0.0247$.  The difference is due to energy
nonconservation.  Grid scale velocity differences are damped by
numerical averaging.  That fraction of the energy extracted from the
shear that does not go into shocks and shock heating winds up in a
turbulent cascade and is lost at the grid scale (one could close the
energy equation by introducing a viscosity, but this introduces a new
set of dynamical complications that will be treated in a separate
publication).  The sense of the difference can be understood by noting
that $\alpha$ measures the rate of energy extraction from the shear
flow.  The extracted energy flows by shock heating into thermal energy
of the disk, and from there to cooling.  In a steady state, all must
balance.  If some of the energy is lost before heating the disk, the
loss must be compensated for by an increase in the rate of energy
extraction.

\subsection{Effect of Cooling Time}

The fates of gravitationally unstable disks can be divided into two
classes, depending on the cooling time.  I find that for short cooling
times, $\tc \lesssim 3 \Omega^{-1}$, the model disk fragments.  This is
easy to understand: for very short cooling times pieces of the disk cool
and collapse before they have an opportunity to collide with one another
and reheat the disk \citep{sb89}.

Fragmentation is illustrated in Figure 5, which shows a snapshot from a
run with $N = 512, L = 80 G\Sigma/\Omega^2, \tc = 2\Omega^{-1}$.  In
this run the disk initially fragments into 6 bound objects, which then
undergo collisional agglomeration.  Ultimately a single bound object
forms which is itself a self-gravitating accretion disk.  This disk has
a rotation frequency $\Omega' \gg \Omega$, and so it is able to sustain
accretion due to repeated gravitational instability because $\tc \gtrsim
3 \Omega'^{-1}$.

When $\tc \gtrsim 3\Omega^{-1}$, the outcome is similar to that of the
standard run: repeated local cycles of cooling, instability, and shock
heating.  According to the analytic theory of \S \ref{ANALYT}, $\alpha =
((9/4) \Omega \tc \gamma (\gamma - 1))^{-1}$ in this regime.  Figure
6 shows $\alpha$ vs. $\tc$ for $\tc > 3\Omega^{-1}$.  The points
are measurements from numerical experiments; the solid line is the
analytic result.  The numerical experiments give slightly high values,
for reasons discussed in \S \ref{STDRUN}.

\subsection{Locality of Angular Momentum Transport}\label{LOCALITY}

The preceding discussion is predicated on the idea that gravitational
instability can be studied in a local model.  It has been argued by
\cite{bp99} (hereafter BP) that the long-range nature of the
gravitational force precludes such a study.  Here I argue in return
that, if certain conditions are fulfilled, it is self-consistent to
study transport of angular momentum by gravitational instability in the
local model.

The local model is an asymptotic expansion of the governing equations
for a thin accretion disk to lowest order in $\eps \equiv
c_s/(r_0\Omega)$ in the neighborhood of a ``fiducial point''.  It
assumes that one is studying structures on lengthscales $\sim \eps r_0$,
that perturbed velocities $\sim \eps r_0 \Omega$, and that the perturbed
potential $\delta\phi \sim \eps^2 r_0^2 \Omega^2$.  The Poisson equation
is solved self-consistently in a WKB expansion (see \cite{shu70} for
higher order corrections).  Even if $\eps \ll 1$ there are several
astrophysically plausible ways the local expansion might fail.

First, structure could exist in the potential on scales large compared
to $\eps r_0$.  For example, a bar might drive a large-scale disturbance
in the neighborhood of the fiducial point.   Bars cannot be generated
self-consistently near the fiducial point because, as I will show below,
gravitational instability only generates structure on scales comparable
to $\eps r_0$.  The bar must be generated elsewhere in the disk, where
$\eps(r) \sim 1$.  It is therefore a condition for validity of the local
model that such large-scale external forcing be absent.

Second, subtler structures may exist in the velocity field, surface
density, or pressure on scales larger than $\eps r_0$.  For example, BP
have argued that gravitational instability can extract energy from
differential rotation, then transform it into waves that propagate over
distances large compared to $\eps r_0$ before dissipating.  Such
long-range coupling would invalidate a local treatment.  While this is a
live and interesting possibility for all types of disk turbulence, BP
use linear theory to argue that self-gravitating disks are qualitatively
different.

Global numerical experiments that contain large scale gradients in disk
properties could falsify or confirm BP's hypothesis regarding wave
transport in self-gravitating disks, but they are beyond the scope of
this paper.  Here I will briefly revisit linear theory to show that
self-gravitating disks are not qualitatively different from
nonself-gravitating disks.  I will then use the outcome of the numerical
experiments to show that long-range correlations in surface density,
which might be expected to develop in the presence of substantial wave
transport, are not present.

Consider a density wave in a razor-thin Keplerian disk.  The disk
structure varies only on a scale $r$, and $\eps(r) \ll 1$.  The wave
$\propto \exp(i (\int dr' k_r(r') + i m\phi - i\omega t))$, where $k_r r
\gg 1$ and  $m /(k_r r) \ll 1$.  The WKB dispersion relation is $(\omega
- m\Omega)^2 \equiv \nu^2 = \Omega^2 - 2 \pi G \Sigma_0 |k_r| + c_s^2
k_r^2$.  The doppler-shifted frequency $\nu^2$ has a minimum at $|k_r|
= k_{cr} \equiv \Pi G \Sigma_0/c_s^2$.  Then the radial energy flux
density, referred to an inertial frame, is
\begin{equation}\label{LINEFLUX}
F_{E,wave} = {1\over{2}} \, \delta \Sigma \, \, c_s^2 \,
	\left( { \nu + m \Omega\over{k_r}} \right) 
	\left({\delta\Sigma\over{\Sigma_0}}\right)
	\left( 1 - {k_{cr}\over{|k_r|}} \right) 
\end{equation}
(\cite{shu70,gt79}).  This is the full wave energy flux.  BP's
``anomalous flux'', by comparison, is the gravitational component of the
energy flux measured in a corotating frame.  Shutting off self-gravity
is equivalent to taking $k_{cr} \rightarrow 0$ in eq.(\ref{LINEFLUX}).
Evidently the wave energy flux does not change qualitatively in this
limit.  

Wave energy fluxes may nonetheless be present.  If they are to change
disk structure significantly, however, they must be of the same order as
the turbulent energy flux $F_{E,wave} \equiv (3/2) \alpha \Sigma c_s^2 r
\Omega$.  If I assume that $\delta \Sigma \sim \Sigma_0$, $Q \sim 1$, and 
$k \sim k_{cr}$, and drop factors of order unity, I find that
$|F_{E,wave}/F_{E,turb}| \sim (m/\alpha) (c_s/(r \Omega))$ (for acoustic
waves in a nonself-gravitating disk, $c_s/(r \Omega)$ is replaced by
$1/(|k_r| r)$).  Thus for 
\begin{equation}
{m\over{\alpha}} \gtrsim {c_s\over{r\Omega}}
\end{equation}
the wave energy flux is as important as the turbulent energy flux.  

To proceed further one can only consider the plausibility of a large
amplitude, high-$m$ wave propagating over significant distances in the
disk.  Here are two arguments against this.  First, a density wave can
only propagate a distance $\sim r/m$ before it turns into an acoustic
wave ($k_r c_s \gtrsim \Omega$).  In a finite thickness disk this
corresponds to a wavelength smaller than a scale height.  If the disk is
stratified three-dimensional effects will modify the wave (e.g.
\cite{ol99}), and the wave is likely to steepen, shock, and dissipate.
Second, the gravito-turbulent state contains fluctuations that emit,
scatter, and absorb waves.  If scattering and absorption are strong, as
they are here, coherent signals are destroyed.  Under these
circumstances it seems unlikely that energy will be transmitted over
large scales by waves.

What can the numerical models tell us about the locality of angular
momentum transport in self-gravitating disks?  I have used two methods
to assess the locality of structure in the nonlinear outcome of my
models.  In the first analysis, I calculate the dimensionless
autocorrelation function of the surface density, $\xi:$
\begin{equation}
\xi(\br) = -1 + {1\over{\<\Sigma\>^2 L^2}}\int \, d^2x' \,
	\Sigma(\br + \br') \Sigma(\br').
\end{equation}
Coherent wave trains would appear as large-scale correlations in the
surface density.  Figures 7 and 8 show the autocorrelation function
averaged from a series of 5 snapshots at $t =
(20,40,60,80,100)\Omega^{-1}$.  Figure 7 shows the spatial structure of
the correlation function from a run with $L = 640 G\Sigma/\Omega^2$ and
$N = 1024$.  Evidently density correlations are concentrated in a region
that is much smaller than the size of the model.  Figure 8 shows cuts
through the correlation function (along the rays marked ``short axis''
and ``long axis'' in Figure 7) that confirm this quantitatively.  

Also shown in Figure 8 is the autocorrelation function for a run with $L
= 320 G\Sigma/\Omega^2$ and $N = 512$ (the same spatial resolution as
the larger model).  Differences between the smaller and larger model
result are small and attributable to sampling noise.  The correlation
function thus appears to depend only weakly on $L$, at least for $L >
320 G \Sigma/\Omega^2$ and $\tc = 10\Omega^{-1}$.  This argues that
surface density structure is locally determined.

In a second analysis, I have calculated which Fourier components of the
surface density dominate the gravitational shear stress.  Figure 9 shows
the quantity
\begin{equation}
{d\alpha_G\over{d k}} = {2\over{3 \<\Sigma c_s^2\>}} \int k d\phi \, 
	{\pi G k_x k_y |\Sigma_{\bk}|^2\over{2 k^3}}.
\end{equation}
Here $\phi$ is an angular coordinate in Fourier space.  This is the
contribution to the gravitational shear stress from Fourier components
of the surface density in the annulus between $k$ and $k + dk$.  The
result is calculated from a model with $L = 640 G\Sigma/\Omega^2$ and $N
= 1024$.  Fully $90\%$ of the angular momentum transport comes from
wavenumbers with $k > 5 (2\pi / L)$. Thus wavelengths significantly
smaller than the model size dominate the shear stress.

Figure 9 can be used to estimate how cool a disk must be for the local
model to be applicable.  If the wavenumber $k_{pk}$ of the maximum in
$d\alpha_G/d k$ is to satisfy $k_{pk} r/(2\pi) \ll 1$, then one needs
\begin{equation}\label{LOCCOND}
{c_s\over{r\Omega}} \ll 0.12.
\end{equation}
where I have used $(2\pi/k_{pk}) \approx 64 G\Sigma/\Omega^2$ and
$\<c_s\> = 7.7 G\Sigma/\Omega$.  YSO disks have $H/r \simeq 0.1$ and so
are at best marginally described by a local approximation.  Some AGN
disk models are much thinner and thus may be accurately described with a
local model.  In disks that violate condition (\ref{LOCCOND}), such as
those studied by \cite{lau96}, global effects can be important.

\subsection{Effect of Resolution}

Most experiments described in this paper have been run at multiple
resolution to determine convergence.  The resolution at which
convergence is achieved depends on what is being measured.  Consider the
convergence of $\alpha$, as measured numerically.  Figure 10 shows
$\alpha$, averaged over the final $80 \Omega^{-1}$ of the run, in a
model with $L = 320 G \Sigma/\Omega^2$ and $N = 32, 64, \ldots, 1024$.
Evidently $\alpha$ has converged.  Investigation of models of different
physical size shows that convergence requires resolution of a fixed
physical scale which is $\approx G\Sigma/\Omega^2$.

\subsection{Length of Run}

The autocorrelation analysis of \S \ref{LOCALITY} shows that large-scale
structures are not established on a dynamical timescale.  Suppose,
however, that gravito-turbulence acts as a viscosity on long-wavelength
modes.  Then there is the possibility of a secular instability analogous
to that of viscous, self-gravitating disks discussed by \cite{lbp74},
and also \cite{gor89,wil92,sch95,gam96b}.  These instabilities grow on
the viscous timescale, $\sim (\lambda / (2\pi H))^2 (\alpha
\Omega)^{-1}$.

As a preliminary test for secular instability, I have integrated two
large ($L = 160 G \Sigma/\Omega^2$), high-resolution ($N = 512$) models
for $10^3\Omega^{-1}$.  One model had $\tc = 40\Omega^{-1}$, the other
$\tc = 10\Omega^{-1}$.  The energy evolution of the latter is shown in
Figure 11.  There is no clear trend in any of the energy components over
the course of the simulation.  The $\tc = 40\Omega^{-1}$ run is similar.
If secular instability is present, it grows only on longer timescales.  

Returning to the analogy with secular instability of viscous,
self-gravitating disks, an integrations over a time $t_f$ can be
regarded as a test for secular instability over wavelengths $\lambda
\lesssim \lambda_c = 2\pi H \sqrt{\alpha t_f \Omega}$.  For the
integrations described here $\lambda_c \approx 30 G\Sigma/\Omega^2$.
Thus longer integrations, although expensive, hold the possibility of
interesting results.

\section{Implications}

What are the implications for the structure of disks around YSOs?
Although our results are not rigorously applicable, since $H/r$ is
typically of order $0.1$--$0.2$ in the outer parts of YSO disks, they
may provide a guide to the relevant physics.

In models of unilluminated YSO disks \citep{rp91,gam96a} $Q$ approaches
unity at a radius $r \sim 30 \au$, where the temperature is $\approx
20\K$.  Illumination can change $Q(r)$ dramatically (\cite{kh87}; for
more recent models see, e.g., \cite{dal99,bel99}).  For sufficiently
strong illumination, or sufficiently low mass disk, the entire disk is
gravitationally stable.  As the illumination is reduced some radii
become gravitationally unstable.  

At gravitationally unstable radii there are two possible outcomes.  If
the cooling time of the disk is sufficiently long ($\tc \gtrsim
3\Omega^{-1}$), then gravitationally driven turbulence will simply
enhance the shear stress above that available from other sources, such
as MHD turbulence.  This shear stress is generated in a region of order
$64 G\Sigma/\Omega^2$ across and all the numerical evidence is
consistent with its being locally determined and treatable as an
$\alpha$ viscosity (but only if $H/r \ll 0.1$).  Recurrent instability
and shock heating maintain $\<Q\> \approx 2.4$, somewhat larger than the
value required for axisymmetric stability.

If, on the other hand, the cooling time in the disk is short ($\Omega
\tc \lesssim 3$), the outcome is dramatically different.  Such short
cooling times might be found in a disk heated mainly by external
illumination that is suddenly switched off (for example, in YSO disks,
as an FU Orionis outburst turns off).  In this case the model disk
fragments, as suggested by \cite{sb89} in the context of AGN disks.  The
long-term evolution of these systems is uncertain because it is
intrinsically global; the fragments undergo collisional agglomeration
within bands in radius.  Clearly one possible outcome is the formation
of gas giant planets (e.g.  \cite{bos97} and references therein), but
the final mass of the object depends on the global evolution.  Global
models including more realistic cooling functions (such as those
developed by \cite{nel00}), and external illumination, are clearly
desirable.

AGN disk models also generally become gravitationally unstable at large
radius \citep{sb89}.  They typically have $H/r \ll 0.1$ and so the local
approximation is more applicable than in YSO disks.  Typically at large
radius their cooling time is short (equivalently: the $\alpha$ required
to prevent instability becomes large).  The numerical models presented
here suggest that fragmentation, and perhaps star formation, is likely
under these circumstances.

I am grateful to Jeremy Goodman, Steve Balbus, Ramesh Narayan, and John
Papaloizou for their comments.  This work was supported by NASA grant
NAG 52837, NAG 58385, and a grant from the University of Illinois
Research Board.

\newpage

% Figure 1.  [t] tells it to go at the top of the next page

\begin{figure}
\plotone{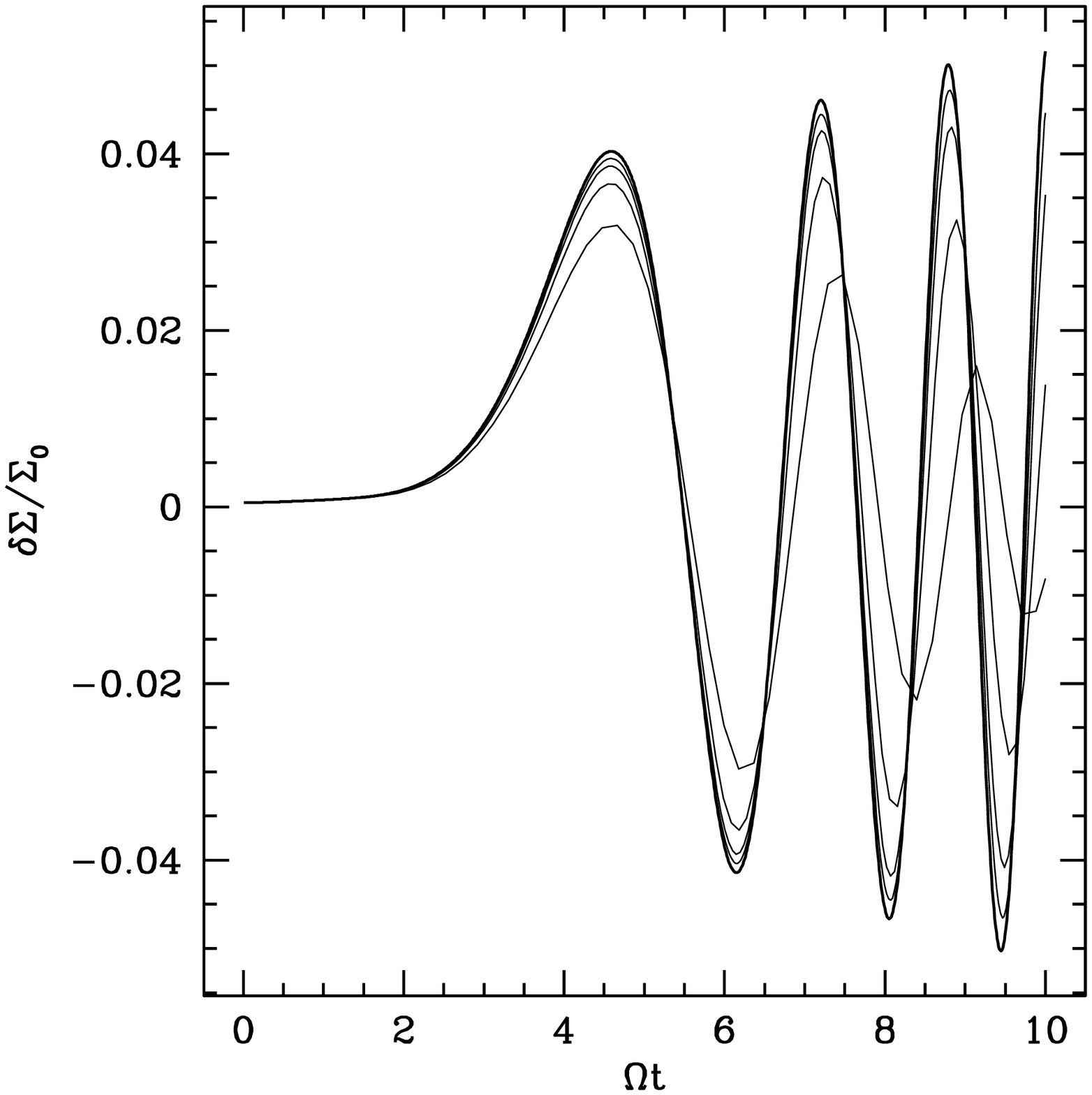}
\caption{
Evolution of the amplitude, in surface density, of a shearing Fourier
component.  The heavy solid line shows the linear theory result.  The
light lines show measurements from a numerical experiment with $N = 32$
(poorest agreement with linear theory), $N = 64, \ldots, 256$.
}
\end{figure}

\begin{figure}
\plotone{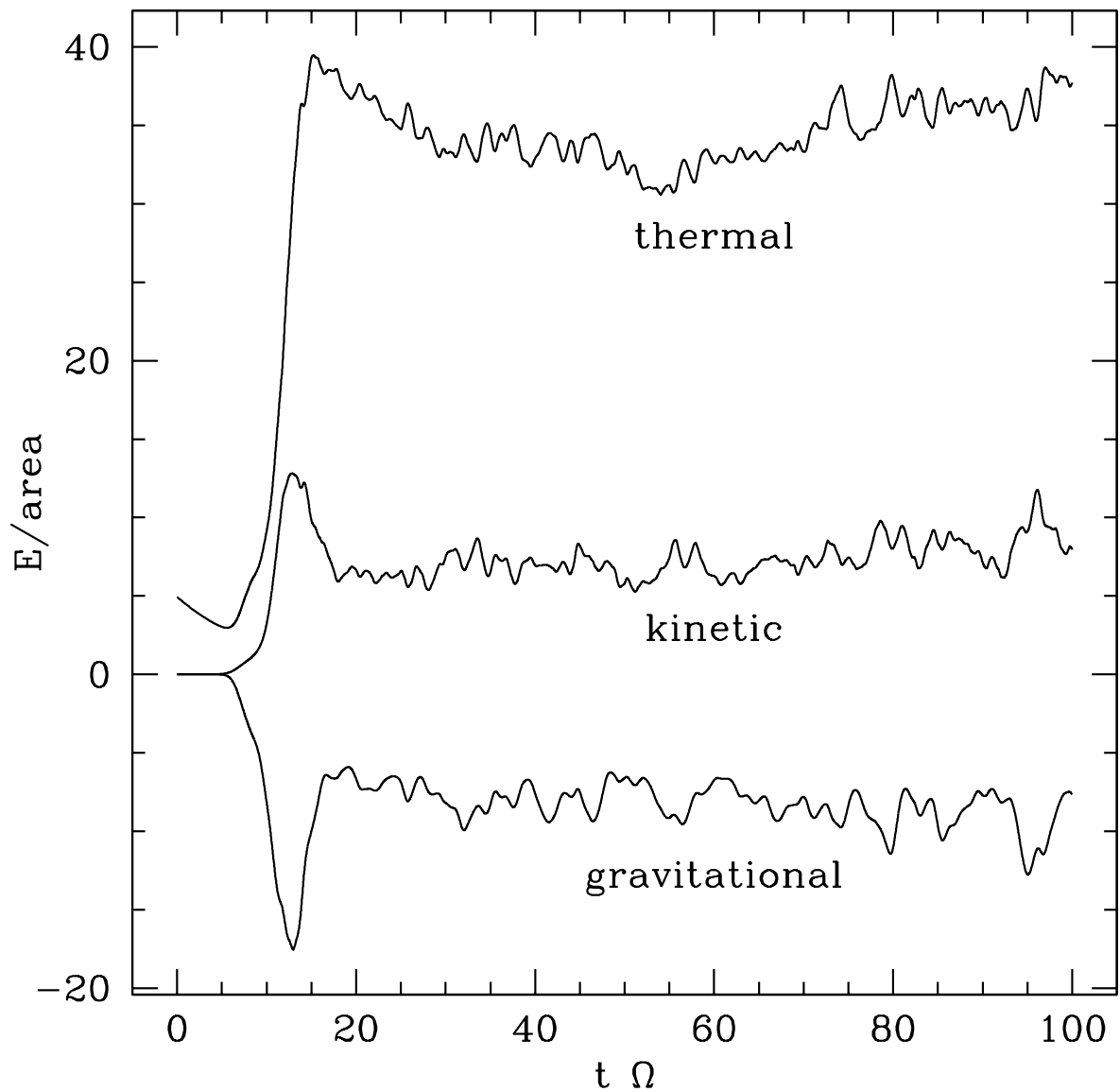}
\caption{
Evolution of the kinetic, gravitational, and thermal energy per unit
area, normalized to $G^2 \Sigma^3/\Omega^2$, in the standard run, which
has $L = 320 G\Sigma/\Omega^2$, $N = 1024$, and $\tc = 10\Omega^{-1}$.
}
\end{figure}

\begin{figure}
\plotone{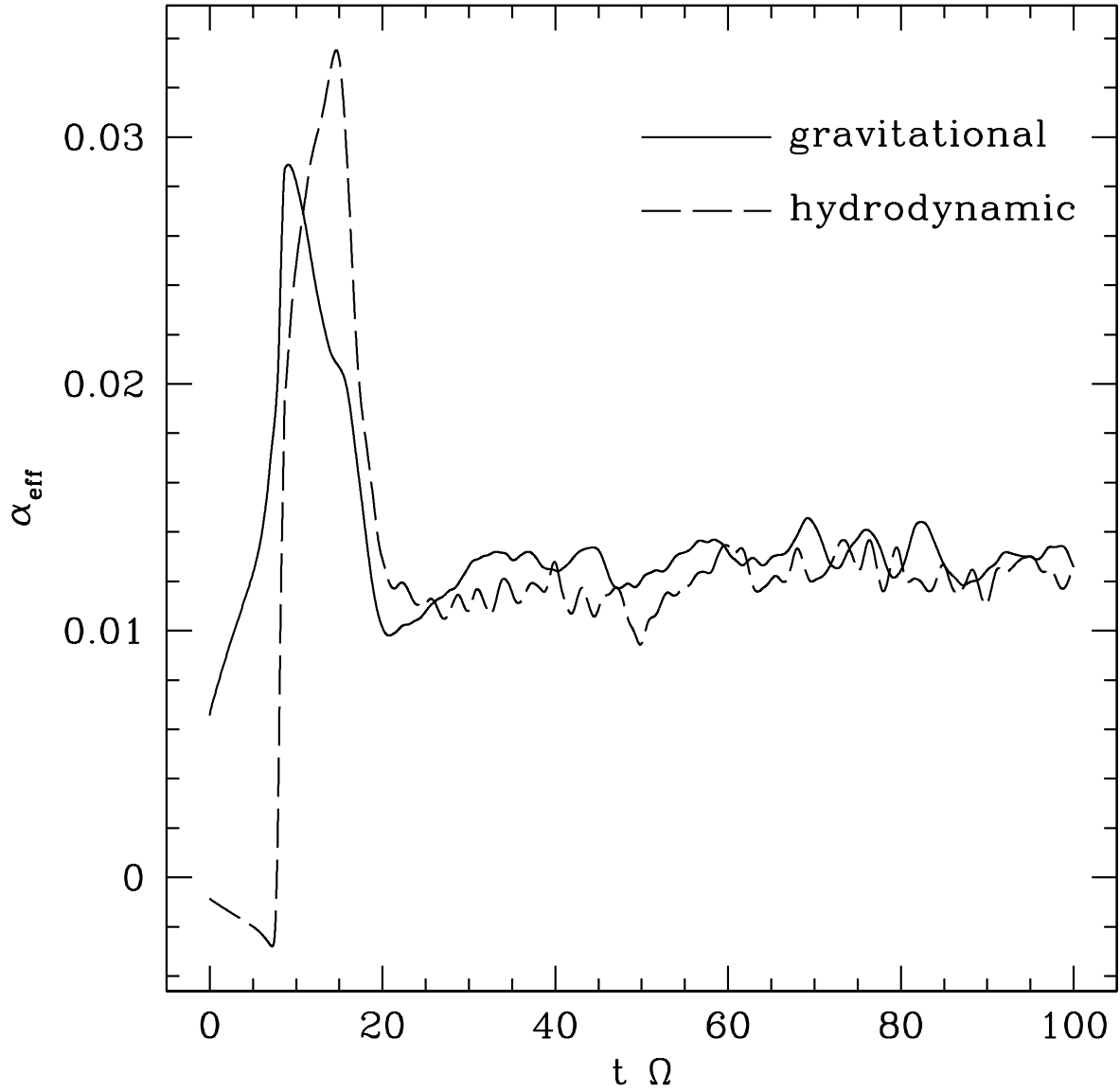}
\caption{
Evolution of the gravitational and hydrodynamic pieces of $\alpha$
in the standard run.  The curves have been boxcar smoothed over an
interval $10 \Omega^{-1}$ to make the plot readable.
}
\end{figure}

\begin{figure}
\plotone{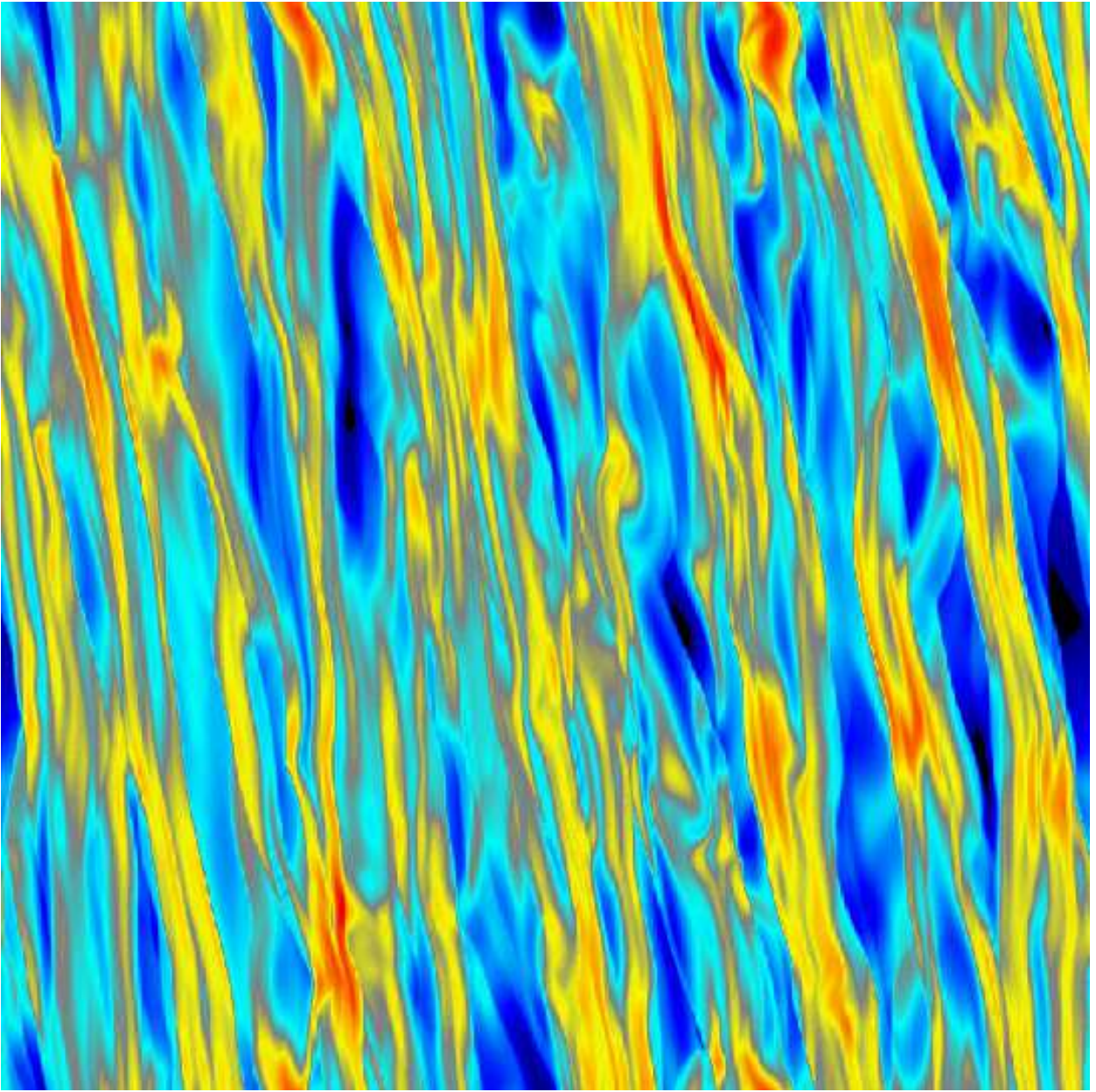}
\caption{
Map of surface density at $t = 50 \Omega^{-1}$ in the standard run.
Black is low density and red is high density.
}
\end{figure}

\begin{figure}
\plotone{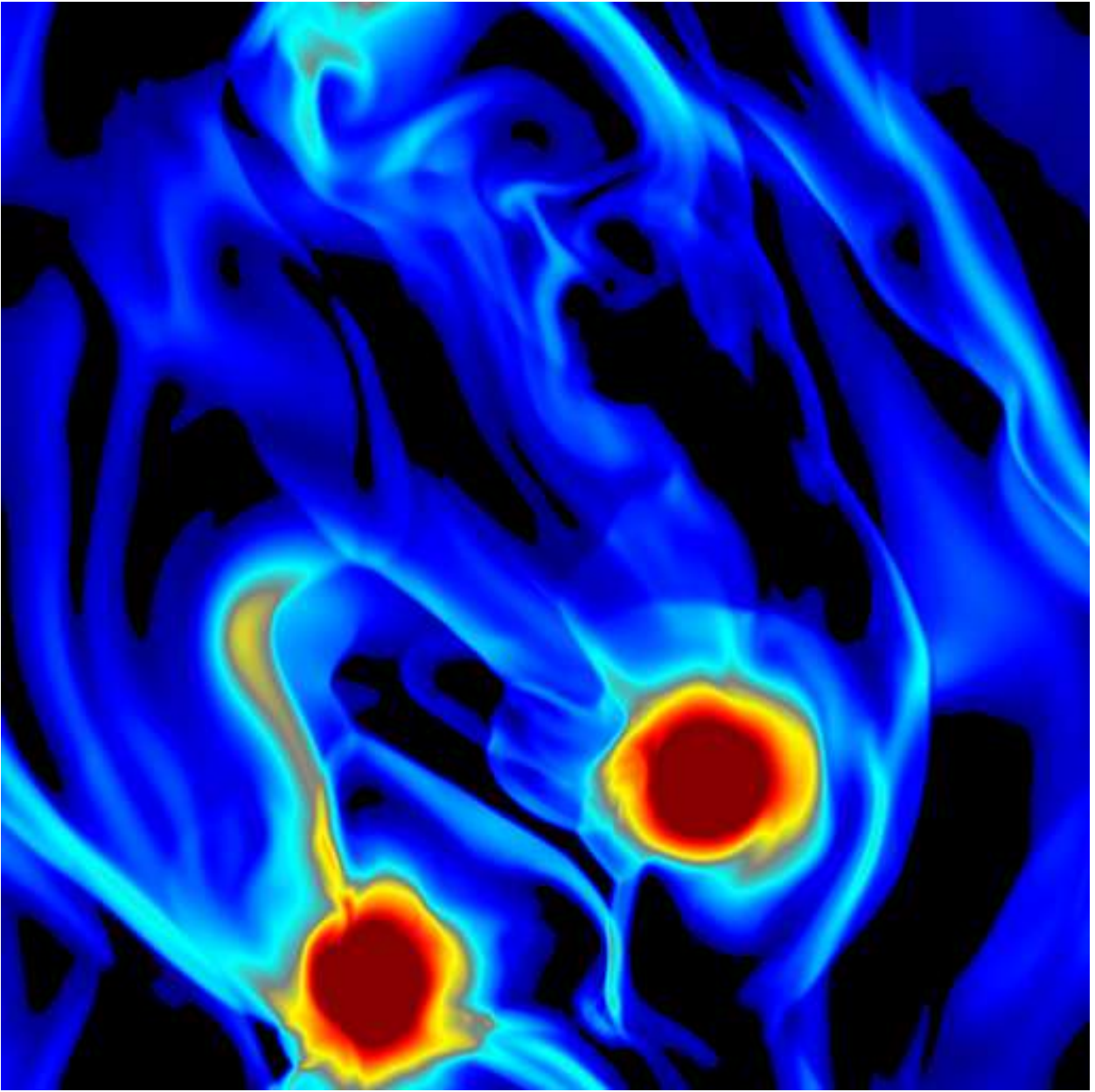}
\caption{
Map of surface density in a run with $\tc = 2\Omega^{-1}$.  Black is
low density and red is high density.  The disk has fragmented and formed
two bound objects.  These objects eventually collide and coalesce.
}
\end{figure}

\begin{figure}
\plotone{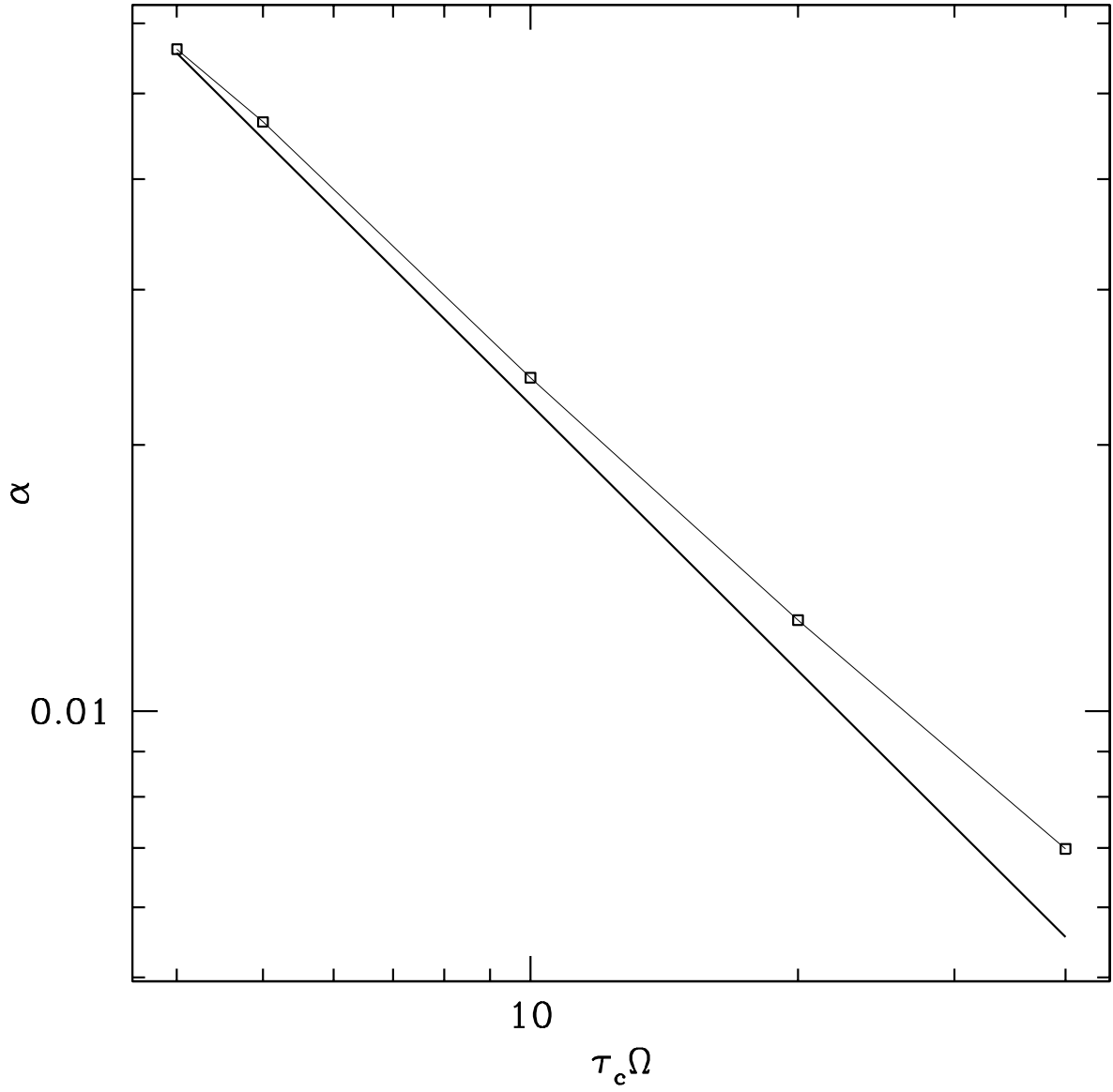}
\caption{
Time-averaged $\alpha$ for a series of runs with $4 \le \tc\Omega \le 40$
(points) and the analytic scaling of \S 3 (solid line).
}
\end{figure}

\begin{figure}
\plotone{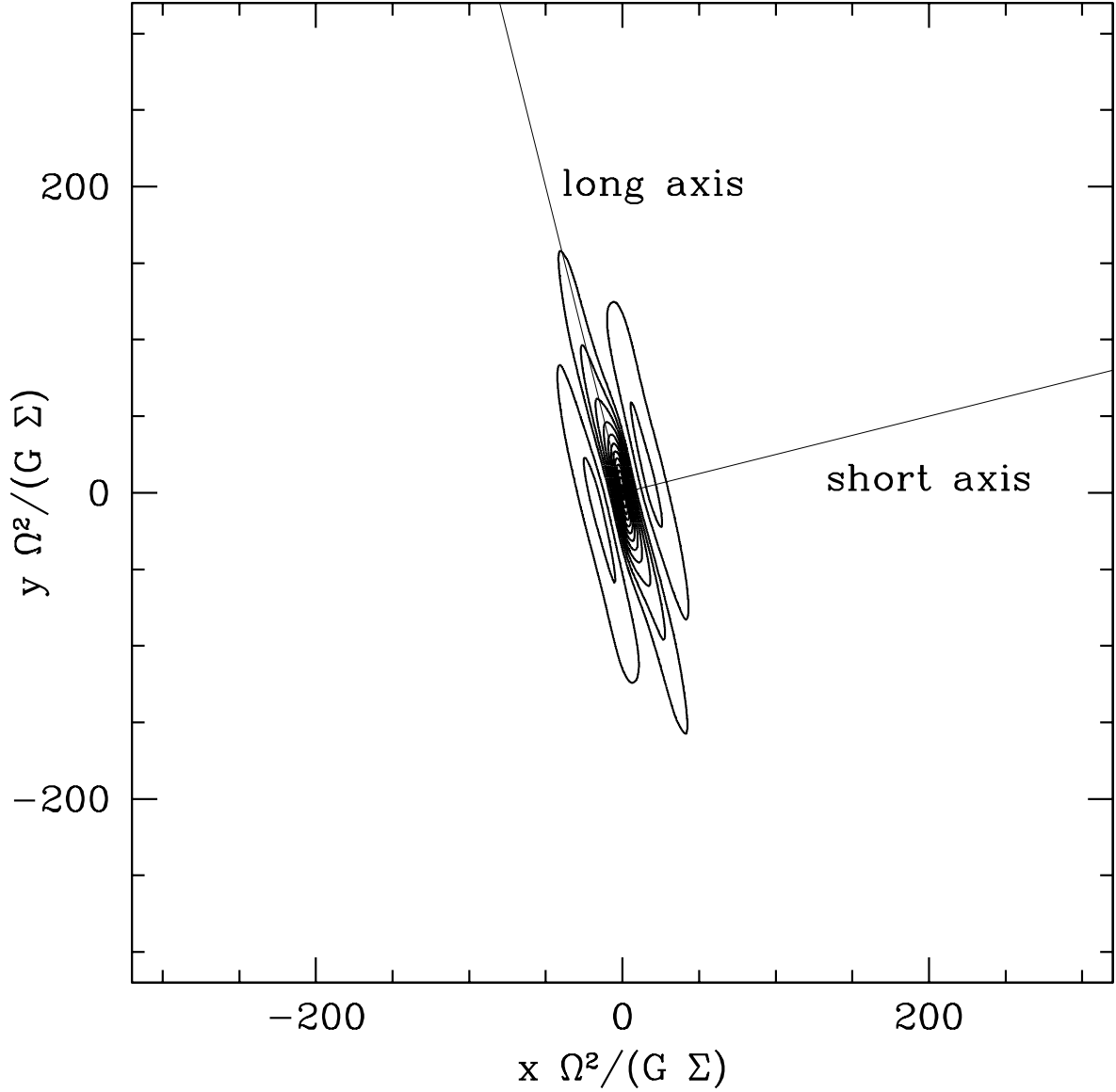}
\caption{
Dimensionless autocorrelation function of the surface density, averaged
over a series of snapshots from a run with $L = 640 G\Sigma/\Omega^2$
and $N = 1024$.  The lines show the cuts along which the correlation
function is sampled for Figure 8.  Contours are
$-0.05,-0.025,0.025,0.05,0.075,\ldots 0.3$.
}
\end{figure}

\begin{figure}
\plotone{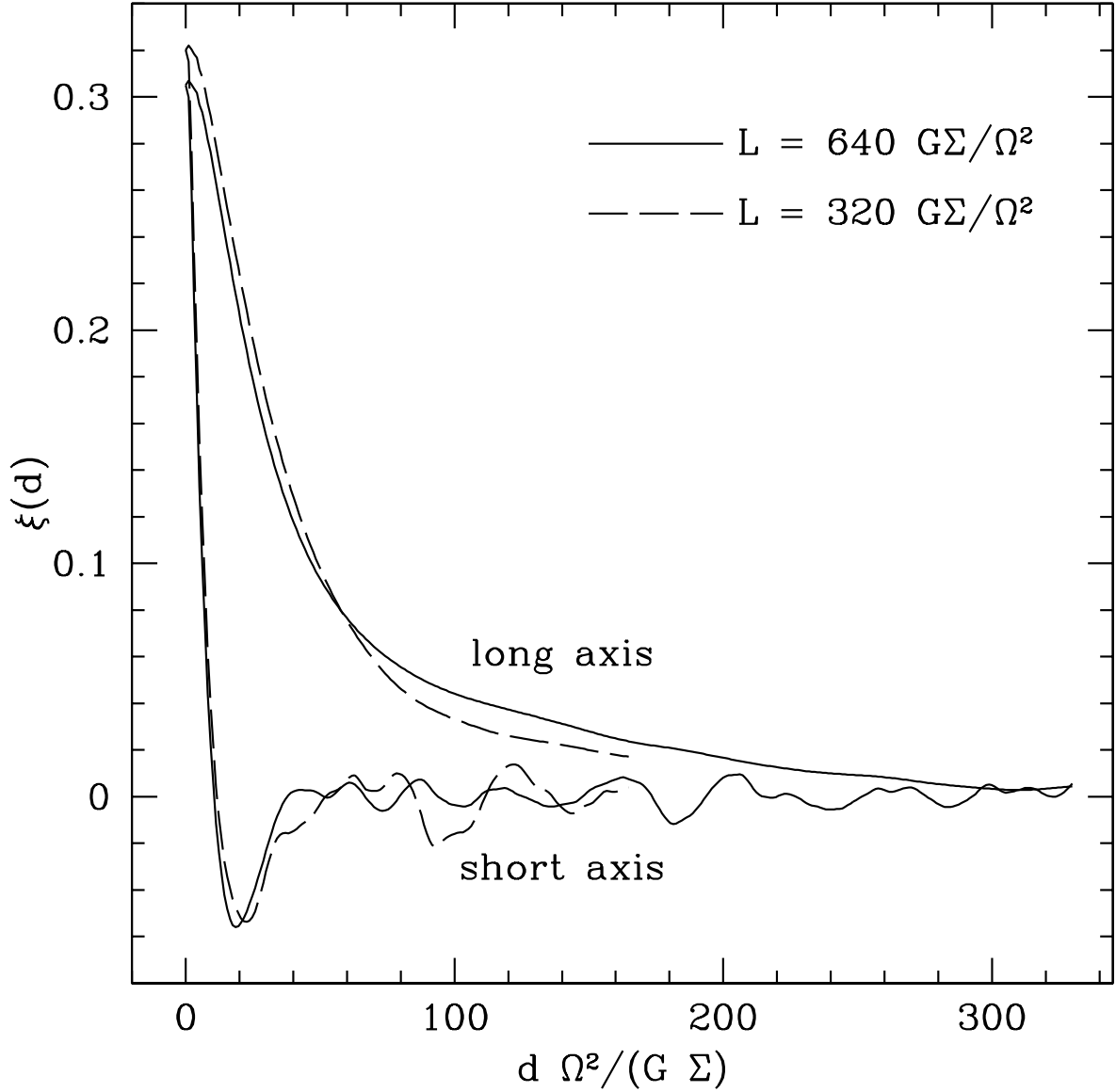}
\caption{
The dimensionless autocorrelation function of the surface density,
averaged over a series of snapshots from a run with $L = 640
G\Sigma/\Omega^2$ and $N = 1024$.  The correlation function is sampled
along the lines shown in Figure 7.  The dashed line is for a run
with $L = 320 G\Sigma/\Omega^2$; the solid line is for a run with
$L = 640 G\Sigma/\Omega^2$.
}
\end{figure}

\begin{figure}
\plotone{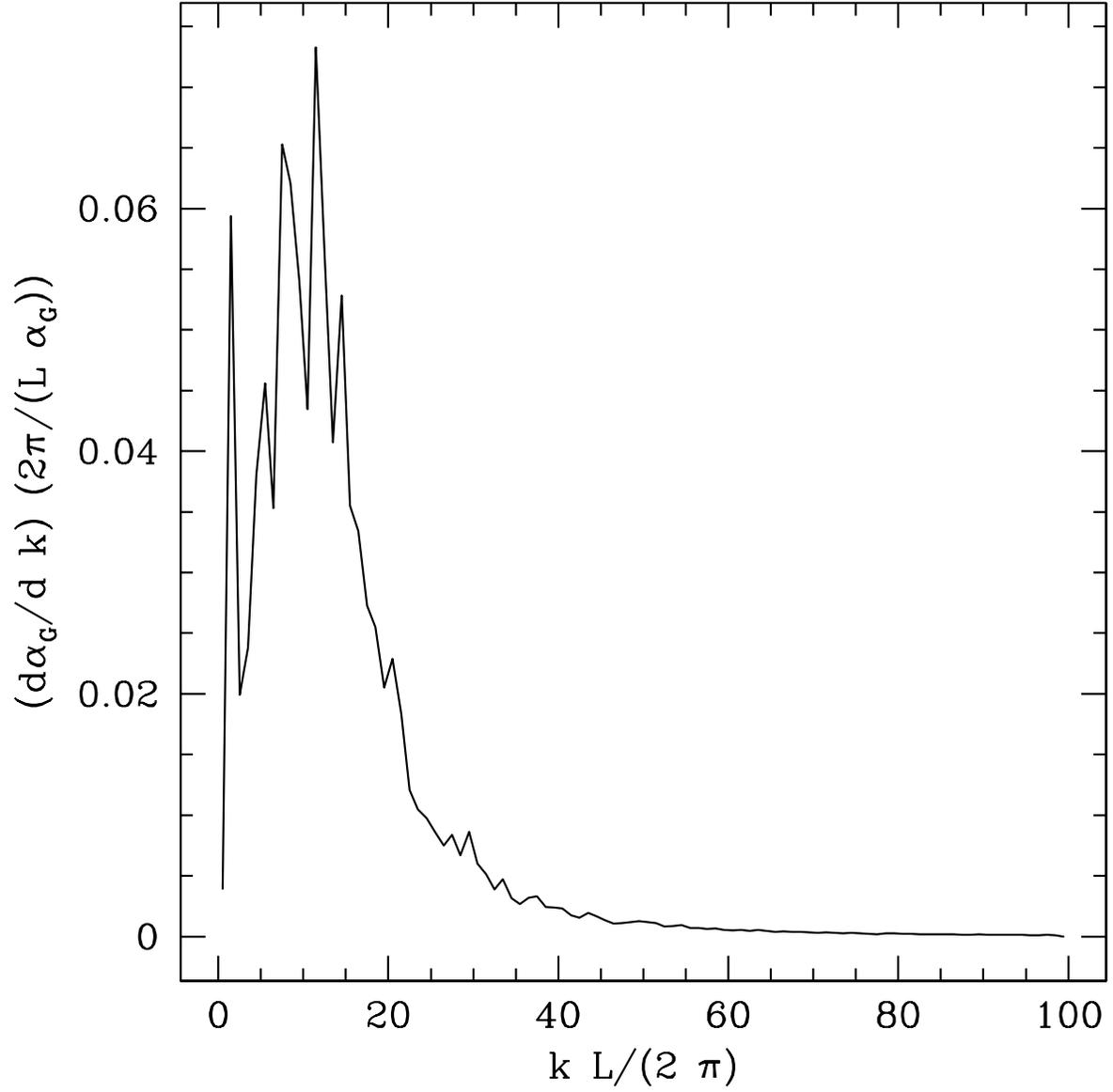}
\caption{
Distribution of the gravitational component of $\alpha$ over $|\bk|$,
from a run with $L = 640 G\Sigma/\Omega^2$ and $N = 1024$.  Fully $90\%$
of the shear stress comes from wavelengths $\lambda < L/5$.
}
\end{figure}

\begin{figure}
\plotone{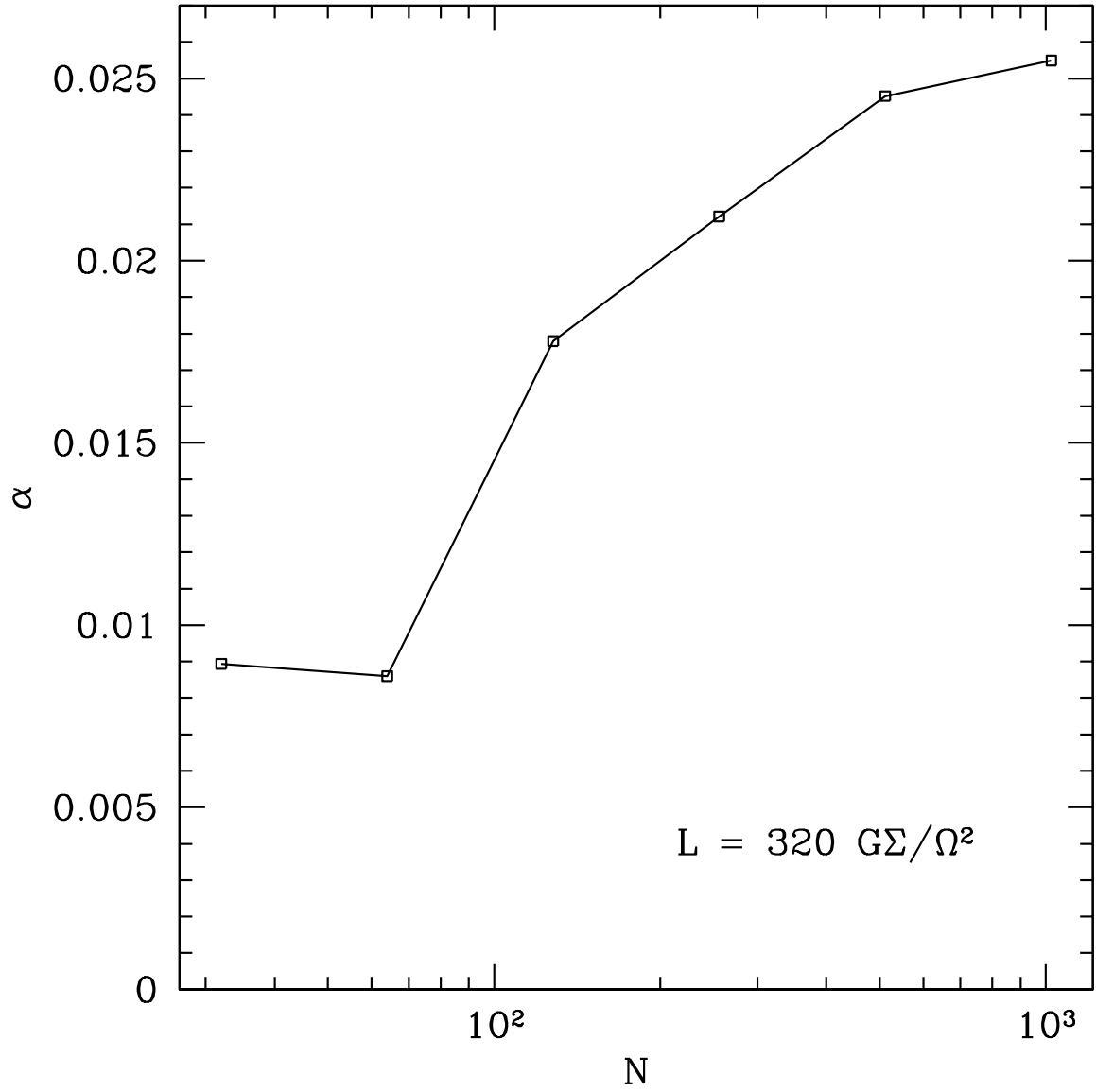}
\caption{
Shear stress $\alpha$ vs. numerical resolution $N$, for
runs with $L = 320 G\Sigma/\Omega^2$ and $\tc = 10\Omega^{-1}$.
}
\end{figure}

\begin{figure}
\plotone{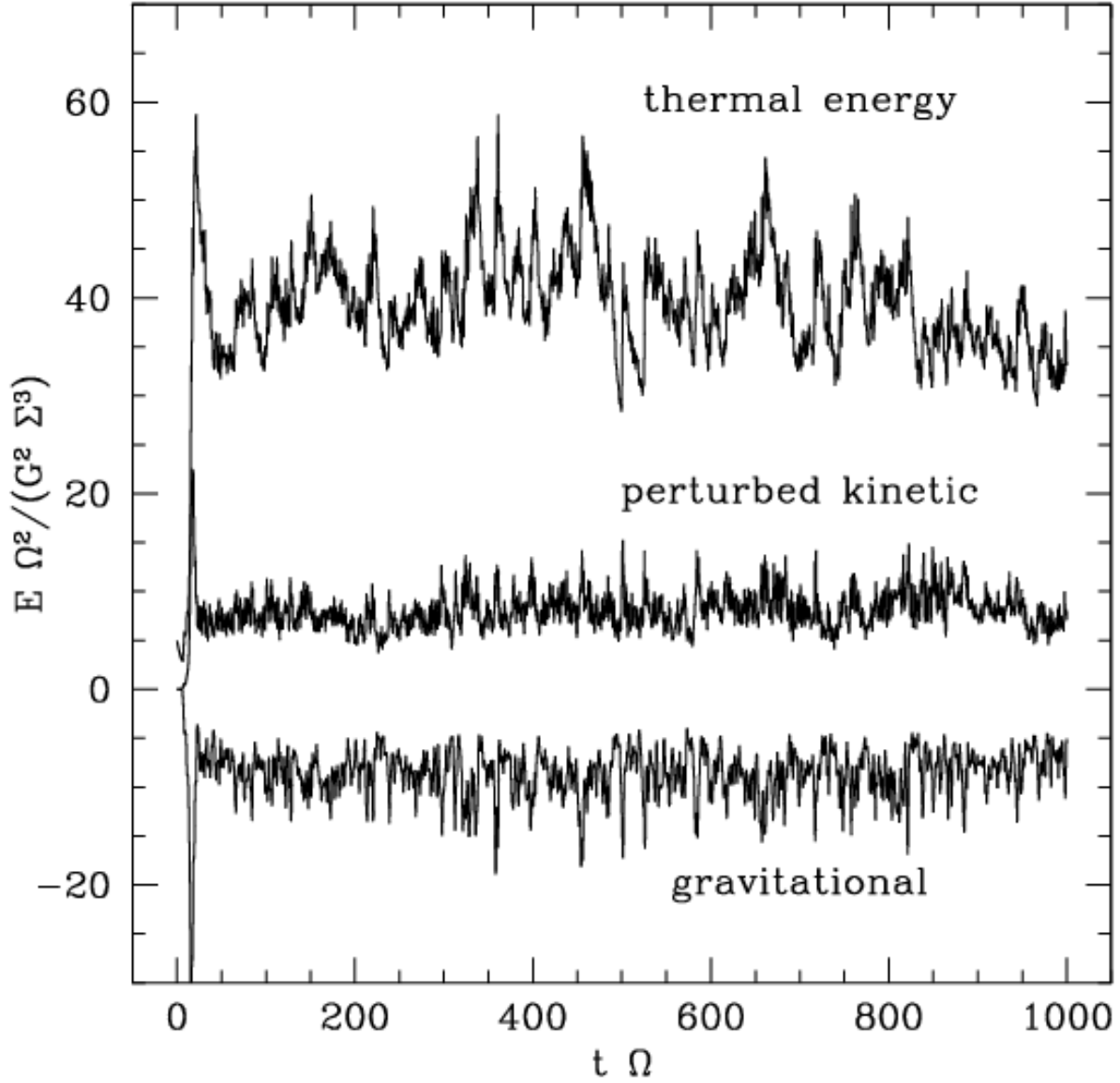}
\caption{
Evolution of thermal, kinetic, and gravitational energy in a long
($10^3\Omega^{-1}$) run, with $\tc = 10\Omega^{-1}$.  No clear trend,
indicative of a secular instability, is present.
}
\end{figure}

\end{document}